# Computation using Noise-based Logic: Efficient String Verification over a Slow Communication Channel

Laszlo B. Kish [1,a] , Sunil Khatri [1] , Tamas Horvath [2,3]

*[1] Texas A&M University, Department of Electrical and Computer Engineering, College Station, TX 77843-3128, USA; email: Laszlo.Kish@ece.tamu.edu ; sunil@ece.tamu.edu*

*[2] Department of Computer Science, University of Bonn, Germany*

*[3] Fraunhofer IAIS, Schloss Birlinghoven, D-53754 Sankt Augustin, Germany; email: tamas.horvath@iais.fraunhofer.de*

**Abstract.** Utilizing the hyperspace of noise-based logic, we show two string verification methods with low communication complexity. One of them is based on continuum noise-based logic. The other one utilizes noise-based logic with random telegraph signals where a mathematical analysis of the error probability is also given. The last operation can also be interpreted as computing universal hash functions with noise-based logic and using them for string comparison. To find out with $10^{-25}$ error probability that two strings with arbitrary length are different (this value is similar to the error probability of an idealistic gate in today's computer) Alice and Bob need to compare only 83 bits of the noise-based hyperspace.

---

[a] Until 1999: L.B. Kiss

# 1. Introduction

In order to demonstrate the applicability of the recently introduced noise-based logic [7-11] to computing, in this paper we describe a case study on using it for the classical problem of randomized verification of string equality over a slow communication channel. As a byproduct of our result, we obtain a noise-based logic realization of a universal hash function as well. To arrive at these results, in this section, we first discuss some recent road-blocks that are being experienced in contemporary Very Large Scale Integrated (VLSI) design. Next, in Section 1.2, we discuss and introduce several deterministic noise-based schemes (Sections 1.2.1 and 1.2.2) which can be employed to overcome the problems being faced in VLSI design. This will form the basis for the following section, which introduces a deterministic noise-based technique for fast comparison of large strings over a slow communication channel.

*1.1 Recent miniaturization trends: choosing limited clock speed to limit bit errors*

Moore's law, which claims an exponential increase of the number of transistors in a microprocessor chip with each successive generation, is arguably in jeopardy as the physical limits to utilize this miniaturization are being approached. The impact of several technical difficulties, such as variability, can be reduced by proper tricks [1] however the limits posed by the laws of physics are firm. In 2002, it was predicted [2,3] that if the trends of those times followed, in 6-8 years, Moore's law will be broken (in terms of increasing the clock frequency and keeping power dissipation under control at acceptable error rates). The prediction was based on the increase and speeding-up of the thermal noise by the shrinking capacitances and increasing noise bandwidth, and the shrinking noise margin due to the required reduction of the supply voltage to keep the gate electrical field and the energy dissipation limited. However, the prediction in [2,3] turned out to be too optimistic because Moore's law was broken one year later, in 2003, when computers with 3 GHz clock frequencies were already on the market with characteristic device sizes of 90 nanometers. As of today (2010), the clock frequency is just slightly above 3 GHz instead of the $\approx 10$ GHz expected from the relevant CMOS transistor bandwidth at the 45 nm device size of today.

One reason for the breaking of Moore's law happening earlier than predicted was the exponentially growing leakage energy dissipation with miniaturization due to the tunneling currents (sub-threshold and gate leakage currents) [4,5]. To control the increased leakage current dissipation requires either to lower the supply voltages with the resulting decrease of noise margin and the exponentially growing error rate, or the reduction of the dynamic energy dissipation by running the processor slower. Apparently, the second choice has been made in the mainstream applications.

Although tricks, such as utilizing bulk bias of the CMOS devices [5] and, at the same time, reducing leakage and dynamical dissipation with giving up speed, the idealistic situation can further be approached. The ultimate 70-80 *kT* energy limit for each switching cycle is an ultimate limit with the expected error probability ($\approx 10^{-25}$) [2,3].

Subsequent studies of quantum informatics [4] and nanoelectronics [6] have indicated that, for general-purpose computing, they are worse alternatives than the CMOS technology because of the much higher error probability and energy dissipation for foreseeable realizations.

*1.2 Some characteristics of the brain and the logic schemes inspired by these*

The limitation mentioned above are motivating the question about how the brain works. Even though the detailed brain logic system is unknown, there are some simple, well-known observations:

i) The number of neurons in the brain is around $10^{11}$, which is a similar number as the number of transistors in a modern computer (including the RAM but excluding solid-state hard drives (a 100G Byte drive has an order of magnitude more transistors).

ii) The neural signal (voltage spikes sequences, "spike trains") in the brain have a maximal spike frequency of about 100 Hz.

iii) These spike trains usually contain spikes fired at random times, often viewed as a quasi-Poissonian process.

iv) The power dissipation of the brain is about 12-20 Watts, which is less than that of typical microprocessors ($\approx 100$ Watts).

*In conclusion, while the brain is using similar energy dissipation and number of active elements as a modern computer (year 2010), it uses at least 30 million times slower signals which are stochastic. These facts raise the question if today's approach with Boolean two-value logic system and deterministic signals is an effective one or rather the brain features are the key to develop more powerful systems?*

Indeed, digital computers are very good at "brute force" applications that require the comparison or computing of a large number of data values, but they perform very poorly with tasks which require intelligence, intuition, etc. There are great problems with even relatively simple pattern recognition tasks such as speech recognition in the presence of background noise, or that of unknown voices.

Recently new deterministic multivalued logic schemes (noise-based logics) have been proposed [7-12] where the information carrier is a system of orthogonal time functions, as discussed below. In most cases, the time functions were stochastic, with continuum amplitude distribution [7,8], random spikes [9,10], or random telegraph signals [11]. As a comparison, however, sinusoidal functions were also tested in chip circuitry [12]. The universality of these logics is proven [7,11].

Here it is important to note that noise-based logic is very different from probabilistic logics, random algorithms, stochastic computing, Brownian-circuits, and the like [13-18].

Noise-based logic [7-11] is deterministic in nature and the role of noise is to carry the information in a multivalued fashion utilizing orthogonality.

*1.2.1 Continuum-noise-based logic and its hyperspace*

Generally, for arbitrary independent stochastic processes with zero mean (reference noises) $V_i(t)$ ($i = 1,...,N$) it holds that:

$$\langle V_i(t)V_j(t)\rangle = \delta_{i,j} \tag{1}$$

where $\langle\ \rangle$ means time average and $\delta_{i,j}$ is the Kronecker symbol (i.e. for $i = j$, $\delta_{i,j} = 1$, otherwise $\delta_{i,j} = 0$). Due to Equation 1, the $V_i(t)$ processes can be represented by orthogonal unit vectors in a multidimensional space. Thus we can use the term *logic basis vectors* for the reference noises, and introduce the notion of an *N-dimensional logic space*, with *logic state vectors* in it [7]. "Deterministic logic" means here that the *idealistic* logic framework is completely *independent* of any notion of probability (there are no probabilistic rules), just like in Boolean logic.

By using this multidimensional space along with linear superposition, logic vectors and their superpositions can be defined, which results in a large number of different logic values, even with a relatively low number *N* of basis vectors. For example, when using binary superposition coefficients that have only on/off possibilities of the reference noises, the number of possible logic values is $2^{2^N}$ in a single wire [8].

Another property of the noise-based logic space is that the product of two different (orthogonal) basis vectors is orthogonal to both the original noises. This property yields a logic hyperspace.

If $i \neq k$ and $H_{i,k}(t) \equiv V_i(t)V_k(t)$ then for all $n = 1,...,N$, $\langle H_{i,k}(t)V_n(t)\rangle = 0$. (2)

$H_{i,k}(t)$ is referred to as a hyperspace vector.

A similar operation can be done with the pairs of hyperspace vectors or with hyperspace vectors and basis vectors, all these with non-overlapping indexes, to grow the hyperspace [7,8], for example:

If $L_{i,k,l,m}(t) \equiv H_{i,k}(t)H_{l,m}(t)$ then $\langle L_{i,k,l,m}(t)V_n(t)\rangle = 0$, $\langle L_{i,k,l,m}(t)H_{p,q}(t)\rangle = 0$ (3)

If $L_{i,k,l}(t) \equiv H_{i,k}(t)V_l(t)$ then $\langle L_{i,k,l,m}(t)V_n(t)\rangle = 0$, $\langle L_{i,k,l,m}(t)H_{p,q}(t)\rangle = 0$ (4)

for every $p,q$, provided $i \neq k \neq l \neq m$. The same type of operations can be continued to generate new hyperspace elements, until we run out of non-overlapping groups of coordinate indexes from the original space. With *N* basis vectors, a $2^N$ dimensional hyperspace can be generated in this manner.

This means that in the cases outlined above, the multiplication operation leads out of the original *N*-dimensional space and introduces higher, new dimensions that are orthogonal to each other and to the basis vectors of the original space.

*1.2.2 Noise-based bits and hyperspace with random telegraph (RTW) signals*

In [11] a new orthogonal multidimensional noise basis for noise-based logic has been introduced: a set of a simple type of random telegraph waves (RTW), $R_i(t_j)$, with discrete amplitudes $\pm 1$, where the index *i* stands for the *i*-th orthogonal basis vector and the index *j* for the *j*-th clock cycle. At the beginning of each clock period, the random telegraph wave takes 1 or -1 amplitude with 50% probability. There is no memory in the system, except that the chosen amplitude is held until the end of the clock period where a new random selection takes place. Henceforth, we drop the continuum time parameter in the notation for an RTW, and refer only to the amplitude of the *i*th RTW during the *j*th clock cycle by $R_i(j)$. The orthogonality of the (infinitely long) $R_i(j)$ and $R_m(j)$ ($i \neq m$, $j = 1, 2, ...$) is the consequence of independently choosing between the amplitudes $\pm 1$ at each clock cycle and each RTW. That means, their cross-correlation coefficient satisfies:

$$\langle R_i(j) R_m(n) \rangle = \delta(i,m) \delta(j,n) \tag{5}$$

Eq. 5 means that whenever the two RTWs are independent or, if they are identical but the time coordinate of one of them is shifted, the cross-correlation coefficient is zero.

The product of an arbitrary number of independently generated RTWs :

$$W_x(j) = \prod_{i=1}^{N} R_i(j) \tag{6}$$

is a hyperspace basis vector which is also an RTW with the same statistical properties and it is orthogonal to the original RTWs or any other RTW generated independently, i.e.:

$$\langle W_x(j) R_k(j) \rangle = \left\langle R_k(j) \prod_{i=1}^{N} R_i(j) \right\rangle = 0 , \tag{7}$$

where either $k \in \{1, ..., N\}$ or $k > N$ meaning that $R_k$ is an RTW generated independently of $R_1, R_2, ..., R_N$.

In the next sections, we show two string verification methods with low communication complexity. One of them is based on continuum noise-based logic. The other one, for which mathematical analysis is also provided, utilizes noise-based logic with RTWs. This operation can also be interpreted as universal hashing with noise-based logic.

## 2. Differentiation between two binary strings via a slow communication channel

Let us suppose Alice and Bob are connected via a slow (or expensive) communication channel. Alice has a binary string $S_A$ of length $L$ and Bob has a binary string $S_B$ of length $M$, where L and M are arbitrary positive integer numbers. They would like to find out with high probability if the strings are different. Below, we show two different noise-based hyperspace schemes to solve this problem with low communication complexity.

*2.1 String verification with continuum-noise based logic hyperspace*

Alice and Bob can make hyperspace basis vectors $W_A(t)$ and $W_B(t)$ representing $S_A$ and $S_B$ by multiplying all the relevant noise bit values:

$$W_A(t) = \prod_{i=1}^{L} S_{A,i}(t) \quad \text{and} \quad W_B(t) = \prod_{i=1}^{M} S_{B,i}(t) \tag{8}$$

where $S_{A,i}(t) = H_i(t)$ or $L_i(t)$, and $S_{B,i}(t) = H_i(t)$ or $L_i(t)$, with respect to the actual bit values, i.e., $H_i(t)$ for high and $L_i(t)$ for low. Note that, in accordance with the principles of noise-based logic, the elements of $\{H_i(t)\}$, $\{L_i(t)\}$ are all zero-mean, independent (orthogonal) continuum noises (see Eq. 1).

If the two strings are identical then the instantaneous amplitudes of the hyperspace vectors will be equal because $S_{A,i}(t) = S_{B,i}(t)$ for all $i = 1,...,L = M$. It is easy to define a low-complexity operation to check this, for example, by constructing and analyzing the difference

$$W_A(t) - W_B(t) = 0 \tag{9}$$

at any moment of time, or the product

$$W_A(t)W_B(t) = \prod_{i=1}^{L=M} S_{A,i}(t) S_{B,i}(t) = \prod_{i=1}^{L=M} S_{A,i}^2(t) \geq 0 \tag{10}$$

of the hyperspace vectors at any moment of time.

Thus a single comparator device checking for non-zero values in the first case or, alternatively, a multiplier and a comparator checking for negative values in the second case can verify if the assumption, that the strings are identical, are violated.

If the strings are different, the generated hyperspace vectors will also be different and deviate a short time into the analysis process. Thus a short sample of $W_A(t)$ sent by Alice to Bob or a short sample of $W_B(t)$ sent by Bob to Alice via the communication channel is enough to detect any difference with high probability.

Note, it is important to clarify the situation of bandwidth. The product of *N* independent noise processes will have *N*-times greater bandwidth, and a corresponding *N*-fold reduction of the correlation time of the resulting signal. Thus, theoretically, an *N*-fold reduction of the correlation time would accelerate the decision of string verification by a factor of *N*. However, this is not practical because our original assumption was a slow channel, which means limited bandwidth. Therefore, at the practical realization of this scheme, Alice and Bob would apply identical low-pass filters, with the original bandwidth, on the product. Thus the string verification would stay as fast as verifying the identity two single noise-bit time functions.

Finally, we must emphasize that, in accordance with Equation 8, the sets of noise bits, $H_i(t)$ for high and $L_i(t)$ for low, are identical at Alice and Bob. This requirement is in accordance with the principles of noise-based logic where all players must have identical sets of reference noises for identification purpose. In the present communication complexity scheme with slow channel, this situation requires either to have a pre-recorded set of reference noises at Alice and Bob, or rather, identical noise generators with identical random number generators and a common, pre-agreed set of random seeds to generate exactly the same time functions for each noise bits in a synchronized way.

*2.2 String verification with RTW based logic hyperspace*

Here we outline the string verification protocol which will then be defined and mathematically analyzed in detail in Section 3. Suppose that $W_A(t)$ and $W_B(t)$ are RTW-based hyperspace vectors where at least one RTW among them is different (difference includes missing RTWs, too). As we have discussed above, their product is again an RTW with the standard statistical properties [11].

Obviously, if the $W_A(t)$ and $W_B(t)$ products contain the same RTW elements, which means that the two strings are identical, the following relations hold:

$$W_A(t_j) - W_B(t_j) = 0 \qquad (11)$$

$$W_A(t_j)W_B(t_j) = 1 \qquad (12)$$

Thus, similarly to the continuum case above, a single comparator device checking for non-zero values in the first case; or a multiplier and a comparator checking for negative values in the second case can verify if the assumption about identical strings have been violated.

If the strings are different, the generated hyperspace vectors will also be different or deviate in a short time. Thus a short sample of $W_A(t)$ sent by Alice to Bob or a short sample of $W_B(t)$ sent by Bob to Alice via the communication channel is enough to detect any difference with high probability.

Suppose that the two strings are not identical. Therefore there will be at least one RTW difference in the products of Alice and Bob. However, even in this case, the first product bit has still 0.5 probability to be identical and that means 0.5 error probability in detecting that the strings are different. The probability that the first two product bits will also be identical is 0.25 which means 0.25 error probability. The probability that first $k$ subsequent product bits are identical, even if the strings are not identical, is $0.5^k$. Thus, if Alice and Bob exchange a $k$-bit long product signal, their error probability in detecting that the strings are different, is $0.5^k$. To reach the theoretical error probability of appr. $10^{-25}$ of the logic gates in computer chips at idealistic conditions, the required $k$ is only 83 (because $0.5^{83} \approx 10^{-25}$). Thus, Alice and Bob are able to detect that there is an arbitrary difference between their bit strings of arbitrary length of $N$ by communicating only 83 bits.

In the next section, the rigorous mathematical proof of these claims and further analysis of the RTW-based string verification is given. Although not discussed explicitly, the analysis presented in the next section can be easily generalized to the continuum noise based string verification problem as well by quantizing the time and the amplitude domain.

**3. Mathematical analysis of the RTW-based hyperspace protocol for verifying string equality**

The string verification by RTW-based noise bits is a simple but powerful application of noise-based logic to the following classical decision problem studied in communication complexity (see, e.g. [19]). Recall from the previous section that Alice has a binary string $S_A$ and Bob has a binary string $S_B$. For the sake of simplicity, but without the limitation of generality, suppose that both strings are of length $L$ and over the alphabet $\{-1,+1\}$ such that Alice doesn't know $S_B$ and Bob doesn't know $S_A$. Their goal is to decide whether or not $S_A = S_B$ by minimizing the communication cost, i.e., by exchanging as few bits as possible. This is a special case of the following problem raised by Yao in his seminal paper [20] on communication complexity: Let

$$f : \{-1,+1\}^n \times \{-1,+1\}^n \to \{-1,+1\} \qquad (13)$$

be some Boolean function and suppose that the full description of $f$ is known for both Alice and Bob. *Given* some binary string $S_A \in \{-1,+1\}^n$ known only for Alice and some binary string $S_B \in \{-1,+1\}^n$ known only for Bob, the *goal* of Alice and Bob is to compute the value $f(S_A, S_B)$ with minimum communication cost. The problem of verifying the equality of two binary strings corresponds to the case that $f$ = EQ and n=L, where

$$\text{EQ}(S_A, S_B) = \begin{cases} +1 & \text{if } S_A = S_B \\ -1 & \text{otherwise} \end{cases} \quad (14)$$

We are interested in a *probabilistic* communication protocol

$$P_\varepsilon : \{-1,+1\}^n \times \{-1,+1\}^n \to \{-1,+1\} \quad (15)$$

that computes EQ correctly with high probability. More precisely, for any error parameter $0 < \varepsilon < 1$ and for any $S_A, S_B \in \{-1,+1\}^n$, the probability of the error must be at most $\varepsilon$, i.e.,

$$\max_{S_A, S_B \in \{-1,+1\}^n} \text{Prob}\big[P_\varepsilon(S_A, S_B) \neq \text{EQ}(S_A, S_B)\big] < \varepsilon \quad . \quad (16)$$

Below we describe such an algorithm in which Alice and Bob use RTWs. The same ideas can be generalized to allow the algorithm to be applicable to the continuum noise case as well. Two main models are distinguished in communication complexity depending on the knowledge shared by Alice and Bob before the computational process starts. In the *private coin* model Alice and Bob use different random binary sequences (i.e., they use different coins for the generation of random binary sequences); in the *common coin* model they use the same random sequences. While the private coin model is much more challenging from a theoretical viewpoint, the common coin model is more realistic. When using RTW based logic hyperspace, (finite) random binary strings correspond to (finite prefixes of) RTWs. Technically, they can be generated, for example, by means of (non-overlapping) sampling of a physical random noise [10]. We note that in the common coin model the exchange of the random binary strings used by Alice and Bob is not charged to the communication complexity, as this happens before the start of the algorithm.

We now describe and analyze the probabilistic protocol in the common coin model for the string equality verification problem outlined in the former section. More precisely, for $0 < \varepsilon < 1$ and positive integer $L$, Alice and Bob generate $2L$ sequences:

$$\begin{aligned} &R_{1,-1}, R_{2,-1}, ..., R_{L,-1} \\ &R_{1,+1}, R_{2,+1}, ..., R_{L,+1} \end{aligned} \quad (17)$$

independently and uniformly at random from the set $\{-1,+1\}^k$ of binary strings of length

$k$, where $k$ is the smallest integer satisfying $k > \log \frac{1}{\varepsilon}$. Each binary string is obtained by taking the prefix of length k of an RTW, or alternatively, by taking $2L$ pairwise non-overlapping infixes of length $k$ of an RTW. These $2L$ random sequences will be fixed for the entire procedure and will be used by both Alice and Bob (common coin). For her sequence $S_A$, Alice first computes the binary string $S_A^*$ of length $k$ defined by

$$S_A^* = R_{1,S_A[1]} \otimes R_{2,S_A[2]} \otimes ... \otimes R_{n,S_A[L]} \tag{18}$$

where $S_A[i]$ denotes the bit value of $S_A$ at the $i$th position and $\otimes$ stands for the component-wise vector product (i.e., $(a_1,a_2,...a_k) \otimes (b_1,b_2,...b_k) = (a_1 b_1, a_2 b_2, ..., a_k b_k)$).

She then transmits $S_A^*$ to Bob, who first computes $S_B^*$ for his string $S_B$ in a similar way, i.e., by

$$S_B^* = R_{1,S_B[1]} \otimes R_{2,S_B[2]} \otimes ... \otimes R_{n,S_B[L]} \tag{19}$$

and then compares $S_A^*$ with $S_B^*$. He concludes that $S_A = S_B$ if $S_A^*$ and $S_B^*$ are equal; otherwise he concludes that $S_A \neq S_B$.

The correctness of this probabilistic protocol follows from

$$\max_{S_A, S_B \in \{-1,+1\}^n} \text{Prob}\left[P_\varepsilon(S_A, S_B) \neq \text{EQ}(S_A, S_B)\right] \tag{20}$$

$$= \max_{S_A, S_B \in \{-1,+1\}^n} \text{Prob}\left[S_A^* = S_B^*\right] \tag{21}$$

$$\leq 2^{-k} \tag{22}$$

$$< \varepsilon \quad , \tag{23}$$

where (23) is immediate from the choice of $k$ and (21) holds by noting that the probabilistic protocol above computes EQ with one-sided error, as it may only fail when $S_A \neq S_B$. To show (22), suppose $S_A \neq S_B$ and let $S_A^{**}$ (resp. $S_B^{**}$) be the binary sequence obtained by taking the component-wise vector product for $R_{i,S_A[i]}$ (resp. $R_{i,S_B[i]}$) for every $i = 1,...,L$ satisfying $S_A[i] \neq S_B[i]$. Since the $R$s have been generated independently and uniformly at random, we have that

$$\text{Prob}\left[S_A^{**}[j] = S_A^{**}[j] \mid S_A \neq S_B\right] = \frac{1}{2} \tag{24}$$

for every $j = 1,...,k$, from which (22) follows by

$$\text{Prob}\left[S_A^* = S_B^* \mid S_A \neq S_B\right] = \text{Prob}\left[S_A^{**} = S_B^{**} \mid S_A \neq S_B\right] \quad . \tag{25}$$

Notice that the number of bits to be communicated is independent of the length $L$ of the input binary strings; it depends only on the error bound $\varepsilon$. Furthermore, the above protocol can easily be adapted to the case that $S_A$ and $S_B$ may have different lengths and Bob (resp. Alice) is not even aware of the length of $S_A$ (resp. $S_B$). The proof above applies to this case as well.

To close this section, we finally note that, from an algorithmic point of view, the probabilistic protocol above is a slight modification of a standard randomized strategy based on *dot products* over the binary Galois field GF(2). (Here GF(2) is the binary field with the elements 0 and 1. Addition is defined by $0+0=0$, $0+1=1$ and $1+1=0$; multiplication is given by $0 \otimes 0 = 0$, $0 \otimes 1 = 0$ and $1 \otimes 1 = 1$.)

One of the differences between the two protocols is that Alice and Bob use only the component-wise vector product in the protocol described above. Though this difference may appear marginal from an algorithmic viewpoint, it is especially important from the point of view of noise-based logic realization of the protocol. As we have shown above, the protocol naturally applies to the hyperspace basis vectors in the two noise-based logic schemes, the continuum and the RTW-based ones, respectively.

### 4. Relationship with universal Hash functions

The proposed string verification techniques are examples of universal hash functions [17,21], but with some key differences. In this section, we first briefly discuss hash functions, and then mention the differences between our RTW and continuum-based string verification approaches and traditional hash functions.

A hash function is a mathematical function which converts elements of a large data set DP (with P elements, say) to elements of a smaller data set DQ (with say Q elements), where Q << P. Each element of DP is mapped to a unique element of DQ. A good hash function is a surjection, with a nearly equal number of DP elements being mapped to any DQ element. If DP corresponds to the set of binary strings of length $L$ and DQ to the set of binary strings of length $k$, where $k << L$ then a universal hash function is a surjection mapping all distinct two elements of DP into DQ independently and uniformly.

Hash functions can be used to speed up membership queries in arrays. In such a case, DQ is an integer, used as an array index. Instead of checking array membership on the set DP, we hash the element being looked up, and search if a secondary array (indexed by the hash function) contains the element being searched. As a result, the membership query complexity reduces from O(|DP|) to O(|DQ|).

In principle, our RTW-based string verification approach is an example of a universal hashing operation. The set DP contains strings $S_A$ and $S_B$, while the set DQ consists of the strings $S_A^*$ and $S_B^*$. The common coin model in this instance is ensured by the fact that both Alice and Bob use the same hash function. The key difference is that a noise based superposition (or multiplication) operation on hyperspace elements is used for determining whether Alice and Bob have a matching string. This demonstrates that both the continuum noise as well as the RTW based approaches can elegantly and efficiently solve the problem of comparing large strings over slow communication channels, with high accuracy.

It is important to note that noise-based logic is more than just a hash function calculator because it is a multivalued universal logic scheme as it has been shown earlier [7-11].

## 6. Conclusion

We showed two string verification methods with low communication complexity where the noise-based hyperspace was utilized. The RTW-based operation can also be interpreted as calculating hash functions. Interestingly, each of the product bits contains a miniscule amount of cumulative information about all the *L* string bits. By using, for example, only 83 bits of the noise-based hyperspace, Alice and Bob can determine with $2^{-83} \approx 10^{-25}$ error probability that two strings with arbitrary length are different. Notice that the error probability of an idealistic gate in today's computer is similar to this value [2,3] which means that communicating more bits for this purpose is meaningless.

These results further strengthen the conjecture that noise-based logic may indeed be utilized by the brain [9, 7]. The particular string verification scheme has some of the common properties of *intelligent decisions*. The key feature of *intelligence* is that it is able to make a decision based on very limited information with a reasonably good error probability.

Let us ignore the complexity need of forming the signal and suppose $L=10^{12}$ and communication speed 1 kilobits/second. In the present scheme the very limited information is the 83 bits communicated between Bob and Alice. Thus Alice and Bob can make an intelligent decision in less than 0.1 second. Using the classical method would mean to communicate all the $10^{12}$ bits and compare the strings bit-by-bit which would take about 30 years.